\documentclass{aa} 
\usepackage{color}
\definecolor{orange}{rgb}{1,0.5,0}
\usepackage{txfonts}
\usepackage{graphicx}
\usepackage{epsfig,latexsym,longtable,lscape,supertabular}
\usepackage[authoryear]{natbib}

\bibpunct{(}{)}{;}{a}{}{,} 
\setcitestyle{authoryear}

\newcommand{\Msun}{M_{\odot}}

\usepackage{txfonts}
\begin{document}
\title{Dark matter halos around isolated ellipticals}

\author{E.~Memola\inst{1,2}, P. Salucci\inst{3}, and A. Babi\'{c}\inst{3}} 

\institute{INFN-Istituto Nazionale di Fisica Nucleare, Sezione di Milano
Bicocca, Piazza della Scienza 3, 20126 Milano, Italy \and INAF-Istituto
Nazionale di Astrofisica, Osservatorio Astronomico di Brera, Via Brera 28,
20121 Milano, Italy \\ email: elisabetta.memola@mib.infn.it \and SISSA/ISAS,
via Bonomea 265, 34136 Trieste,  Italy } \abstract
{}
{We investigate the distribution of the luminous and the dark matter components
in the isolated ellipticals NGC~7052 and NGC~7785, which are embedded in an
emitting hot gas halo, by means of relevant X-ray and photometric data.}
{To calculate the dark matter distribution in these rare objects, 
we performed an improved X-ray analysis of the XMM-Newton data of NGC~7785, 
and we used former results based on Chandra data of NGC~7052. 
For each object we also derived the stellar spheroid length scale from 
the surface photometry and
the spheroid stellar mass from an analysis of the galaxy 
spectral energy distribution.}
{We find that a dark matter component is present in these objects. It is
subdominant and mixed with the luminous matter inside the optical region
half-light radius wide, while it dominates the gravitational
potential at outer radii. On the whole, the dark halo structure is very
similar to that found around spirals of comparable luminosity and it is well
reproduced by a Burkert halo, while a S\'{e}rsic spheroid accounts well 
for the baryonic component.}
{}

\authorrunning{E.~Memola et al.}
\titlerunning{Dark matter in isolated ellipticals}

\maketitle

\section{Introduction}

\setcitestyle{authoryear} The dynamics of galaxies is influenced, and
often dominated, by non-radiating matter, which reveals itself only
through a gravitational interaction with the luminous matter. From the
study of a large number of disk systems, a one-to-one relation between
the (luminous) central object and the massive dark halo that envelopes
it has emerged \citep[see][]{shankar,mandelbaum09}. The paradigm is that every
galaxy lies within a self-gravitating dark halo of size $R_\mathrm{vir}$
and mass $M_\mathrm{vir}$: more specifically, disks of sizes
$R_\mathrm{opt}$ and luminous masses $M_\mathrm{\star}$ are embedded in
dark halos of virial radii $R_\mathrm{vir} \sim 15 \, R_\mathrm{opt}$
and masses $M_\mathrm{vir} =3 \times 10^{12} \left(\frac{M_*}{2\times
10^{11}\Msun }\right)^{0.4}\Msun$, where $\Msun$ is the solar mass 
\citep{persic96,kormendy,salucci07,oh08,donato}.

Do ellipticals follow a similar scenario
\citep[see][]{bertola,salucci,tortora}? 
The answer would be of crucial importance for understanding
how galaxies form and perhaps even for the nature of the dark
matter (DM) itself. Unfortunately, evidence on the properties of
the mass distribution in these objects is presently not conclusive.

Indeed, to derive the mass distribution from stellar kinematics, 
as is done in spirals where
rotation balances gravity, is very challenging.
Ellipticals are pressure-supported stellar systems whose
orbital structure may also involve some angular momentum content,
a degree of triaxiality, and velocity dispersion anisotropies.
As is well known, the derivation of the mass
distribution from the stellar motions in these systems
through Jeans equation
and its higher moments is quite complex. 
It is difficult to obtain
the galaxy gravitational potential unambiguously and, 
following this, the mass distribution. 
Furthermore, in these
systems the luminous matter is very concentrated toward the
center, where it dominates the gravitational potential. Therefore,
the motions of the main baryonic component in ellipticals are
generally a limited tracer of the distribution of the dark matter,
which mostly lies outside the stellar spheroid (e.g., Kronawitter
et al.~2000). Only recent dynamical measurements
reached the DM dominated regions 
(e.g., Thomas et al.~2009 and references therein; Pu et al.~2010).

Alternative mass tracers (planetary nebulae, ionized globular clusters,
and neutral disks) from baryonic components of negligible mass are
often a better probe of the total gravitational potential \citep[see][and
references therein]{roman}. In particular, the X-ray emitting hot gas
is known to extend out to very large radii and to trace the
gravitational potential.
This method yields accurate mass profiles \citep[e.g.,][]{forman,osully07,zhang} 
and it is very robust because it provides reliable mass profiles also for highly
disturbed objects where the hydrostatic equilibrium is far from being
established \citep{johnson}. 

This method works best for isolated systems and has the rare advantage
that it can probe the gravitational potential from almost the galaxy
center out to the halo virial radius. Moreover, to obtain mass
profiles of objects that never resided in a cluster or group resolves the
question of whether ellipticals possess their own DM halos, or 
``build them'' from the cluster DM, and it allows us to
investigate the environmental dependence of the DM distribution in galaxies.

\section{X-ray emitting halo ellipticals}
A regular X-ray morphology is commonly assumed to indicate approximate 
hydrostatic equilibrium, and the absence of nearby companions leaves the 
X-ray halo unperturbed.
The gravitating mass inside a radius $r$, $M(r)$, can be
derived from the X-ray flux under the assumption that the emitting gas is in
hydrostatic equilibrium. As is well known, from the gas density and
temperature profiles \citep[e.g.,][]{fabricant} we obtain:
\begin{equation} 
 M(<r) = {kT_\mathrm{g}(r) r \over { G \mu m_\mathrm{p}} }\Big({d\log\
 {\rho_\mathrm{g}}\over{d\log\ r}} + {d\log \ T_\mathrm{g}(r)\over{d\log \ r}}\Big),
 \label{rybi} 
\end{equation}
where $T_{\rm g}$ is the gas temperature at radius $r$, $\rho_\mathrm{g}$ is the gas
density, $k$ is the Boltzmann constant, $G$ is the gravitational constant, $\mu
$ is the mean molecular weight that we set at the value of $\mu=0.62$
\citep{ettori}, and $m_\mathrm{p}$ is the mass of the proton.

The observed X-ray surface brightness, $\Sigma_\mathrm{g}$ is usually fitted by a $\beta$-model (Cavaliere \& Fusco-Femiano 1976):
\begin{equation}
\Sigma_\mathrm{g} \propto [1 + (r/r_\mathrm{c})^2]^{-\beta/2}
\label{eq2} 
\end{equation}
that yields
\begin{equation}
\rho_\mathrm{g} \propto [1 + (r/r_\mathrm{c})^2]^{-3 \beta/2},
\label{eq3} 
\end{equation}
with the core radius $r_\mathrm{c}$, and the slope $\beta$ as free parameters. 
For an isothermal gas distribution: 
$
dlog T_g(r)/ dlog r = 0,
$
therefore,
we have, in a spherical case, 
from the above equations:
\begin{equation}
M(r) \propto {(r/r_\mathrm{c})^3 \over {[1 + (r/r_\mathrm{c})^2]}}.
\label{mr}
\end{equation} 
%
An isothermal distribution is often found in ellipticals; however, there 
are cases in which the temperature has a profile that must be considered 
in Eq.~(1), although its contribution is almost always much smaller than 
that from the gas density gradient (e.g.~\citealt{fukazawa}). 
In the present work, we have a very low spatial resolution 
temperature measurement, therefore we are forced to assume that the 
temperature in the region under analysis, varies linearly with radius. 
An assumption that, however, agrees well with most 
of the high-resolution temperature profiles.
Then, Eq.~(4) or, when the temperature gradient is different from zero, Eq.~(1)
yields the gravitating mass as a function of the radius in the region where the
X-ray halo emits.

\begin{figure*}
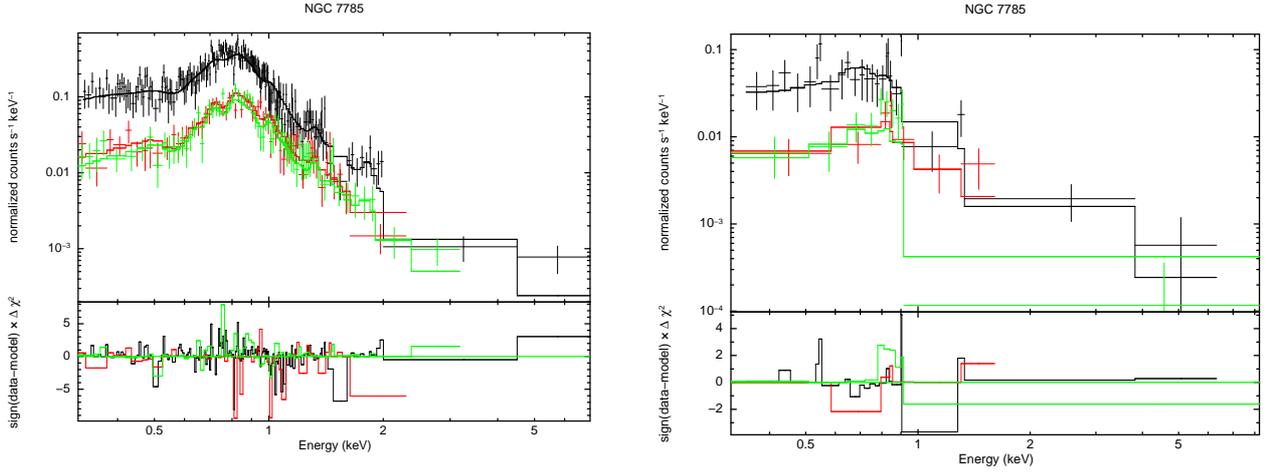

\resizebox{17cm}{!}{
\hspace*{1cm}
\psfig{figure=tris_65_froz.ps,width=15.5cm,angle=-90}
\hspace*{2cm}
\psfig{figure=tris_65-130_froz.ps,width=16.0cm,angle=-90}
}
\caption{
XMM-{\em Newton} EPIC-MOS1 (red), -MOS2 (green), and -pn (black) spectrum and
$\chi^2$ behavior for NGC~7785. {\em Left panel}: extraction radius of
65\arcsec; {\em Right panel}: extraction annulus with 65\arcsec and 130\arcsec
radii. The best-fit model (solid line) is a power-law plus a thermal component (see color
figure online).
\label{7785spec}}
\end{figure*}

\section{ The isolated ellipticals NGC~7052 and NGC~7785}
The two elliptical galaxies we study here are members of a sample of 43
bright isolated galaxies \citep{focardi11} selected from the Updated
Zwicky Catalog \citep[UZC,][]{falco} by means of an adapted version of
the \citet{focardi} neighbor-search code. Seven of those are early-type
objects, for four of them results of the X-ray analysis of proprietary
and archival data have been published by \citet{memola}, and
finally, for two of them (\object{NGC~7052} and \object{NGC~7785}) the
available X-ray statistics allows us to derive the dark matter content
and distribution. 

We recall the criteria to select the sample of {\em truly}
isolated galaxies: {\it a)} absolute B luminosity higher than $1.3 \times
10^{10}h_{75}^{-2}$ L$_{\rm B\odot}$, where L$_{\rm B_\odot}$ is the $B$-band
solar luminosity; {\it b)} recessional velocity in the range [2500-5000]
km s$^{-1}$; {\it c)} galactic latitude ($b^{\rm II| }\ge 15\degr $); and
{\it d)} no companion galaxies down to 2 magnitudes fainter than the target
galaxy within a circular radius of $1.3 h_{75}^{-1}$~Mpc and at velocity
distance of 1000~km~s$^{-1}$. These ellipticals are selected in an
objective way and their isolation is granted at a typical cluster/group
scale.  Moreover, NGC~7785 is present in the AMIGA database of isolated
galaxies \citep{verdes} and is also part of the \citet{Smith} sample. 


The importance of selecting objects in total isolation was discussed in the
introduction: dark matter halos of {\em cluster ellipticals} merge with the
cluster halo at a scale of $\sim 300\ $kpc, a scale comparable with that of
their virial radii.  This complicates the dynamical analysis, and more
importantly, it implies that the present-day halos around cluster galaxies are
not the cosmological dark matter halos anymore, indeed, ellipticals at
the center of clusters are found to have more DM than ellipticals in 
low-density regions \citep{nagino}. 
The very low density of the environment of isolated ellipticals makes them 
precious laboratories for investigating the cosmological properties of DM halos 
around early-type galaxies.

Note that NGC~7052 and NGC~7785, 
with L$_X<2\times10^{41}$ erg\,s$^{-1}$, and M$_{\rm TOT}<2 \times 10^{12}M_{\odot}$ 
(see Table 3 in Memola et al.~2009) do not appear to be {\em fossil groups} 
(e.g.,~Jones at al.~2003; Pompei et al.~2007) because they do not show
their physical properties. Fossil groups 
(e.g., NGC 1132, Mulchaey\&Zabludoff 1999)
have, indeed, so far been observed with 
larger masses $10^{13}-10^{14}M_{\odot}$, 
and X-ray luminosity $>10^{42}$ erg\,s$^{-1}$ 
(see Fig.~8 in Memola et al.~2009).

\section{X-ray observations}
\subsection{X-ray data}

NGC~7052 and NGC~7785 have extended X-ray halos out to 16 and 32~kpc,
respectively, with hot gaseous mass of $M_\mathrm{gas}=2.2 \times 10^9
\Msun$, and $M_\mathrm{gas}=4.6 \times 10^9 \Msun$. {\em Chandra} observations
of NGC~7052 and XMM-{\em Newton} observations of NGC~7785 have been presented
and discussed in \citet{memola}. NGC~7052 was observed by the {\em Chandra}
ACIS-S instrument in September 2002 for about 10~ks, and NGC~7785 was observed
by the XMM-{\em Newton} EPIC detectors in June 2004 for about 15~ks
(PrimeFullWIndow mode and thin filter). For the spectral analysis of NGC~7052
we refer to \citet{memola}.

The net radial surface brightness profiles, centered at the X-ray peak of
the EPIC-pn data in the 0.5-2.0 keV energy band, are well reproduced by a
$\beta$-model whose best-fit values of $r_{\rm c}$ and $\beta$ are given in
Table~\ref{par-tab} \citep[see also][]{memola}.

\begin{table}
\caption{Physical properties of the galaxies. 
Columns:
(1) galaxy name;
(2) core radius of the best-fit $\beta$-model; 
(3) slope $\beta$;
(4) absolute $B$-band magnitude; 
(5) $B$-band magnitude;
(6) $K_\mathrm{s}$-band magnitude, 2MASS total magnitude;
(7) scale in kpc/$\arcsec$, which also provides the distance assumed. 
$B$- and $K_\mathrm{s}$-band magnitudes are from LEDA or NED.}

\begin{tabular}{ccccccc}
\hline 
\hline
\noalign{\smallskip}
Source & r$_{\rm c}$ & $\beta$ & M$_{\rm B}$ & m$_{\rm B_{TC}}$ & K & kpc/$\arcsec$ \\
(1)& (2)& (3)& (4)& (5)& (6) & (7) \\
\hline
\noalign{\smallskip}
NGC~7052 & 1.11'' & 0.48 & -21.46 & 12.73 & 8.57 & 0.323 \\
NGC~7785 & 35'' & 0.95 & -21.38 & 12.34 & 8.45 & 0.247 \\
\noalign{\smallskip}
\hline
\label{par-tab}
\end{tabular}
\end{table}

\begin{table*}
\caption{ NGC~7785: summary of the XMM-{\em Newton} spectral results. The EPIC-pn net
counts are derived from the regions (a circle and an annulus) of radii given in the second
column.}
\label{7785tab}
\begin{center}
\begin{tabular}{cccccccc}
\hline 
\hline
\noalign{\smallskip}
N$_{\rm H_{gal}}$ & region$_{\rm ext}$ 
 & cts & kT$_{\rm mekal}$ & $\Gamma$ & $\chi^{2}/dof$ & L$_{\rm mekal
 (0.5-2.0)}$ & L$_{\rm po(2.0-10)}$ \\
cm$^{-2}$ & $\arcsec$ & pn & keV & fixed & & erg s$^{-1}$ & erg s$^{-1}$ \\
\noalign{\smallskip}
\hline
\noalign{\smallskip}
5.24 $\times 10^{20}$ & 65 & 2687$\pm$59 & 0.57$^{+0.01}_{-0.02}$ & 2.0 & $307/314$ & 6.3 $\times 10^{40}$ & 6.9 $\times 10^{39}$ \\ \noalign{\smallskip}
5.24 $\times 10^{20}$ & 65--130 & 615$\pm$64 & 0.32$^{+0.09}_{-0.05}$ & 2.0 & $39/39$ & 9.4 $\times 10^{39}$ & 5.4 $\times 10^{39}$ \\
\noalign{\smallskip}
\hline
\end{tabular}
\end{center}
\end{table*}

\begin{figure*}
\resizebox{18cm}{!}{
\hspace*{0.5cm}
\psfig{figure=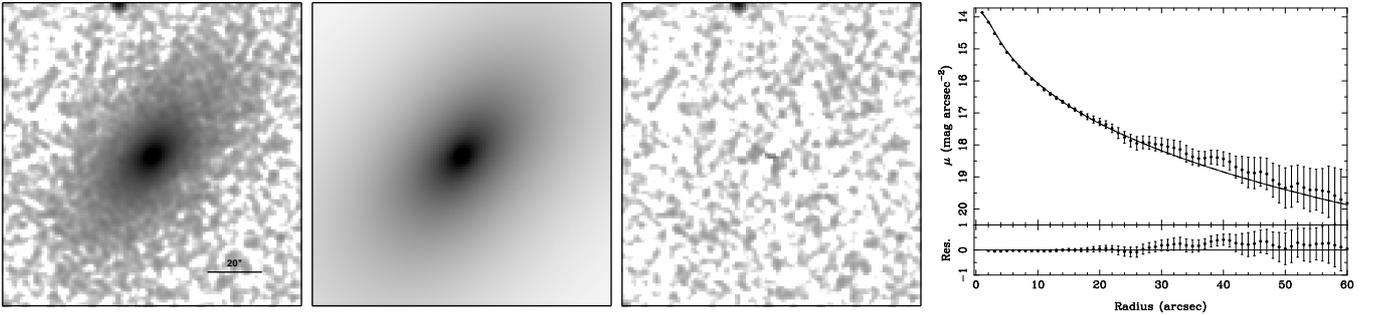,height=14.5cm}
\hspace*{0.5cm}
\raisebox{14.2cm}{\psfig{figure=01-ngc7785-aK-lin.ps,width=14.5cm, height=19.6cm, angle=-90}}
}
\caption{ 
Two-dimensional 2MASS $K_\mathrm{s}$-band image decomposition of NGC~7785. From left
to right: the original image, the model image, the
residual image, and the surface brightness profile plot. The image surface
brightness profile is marked by filled circles and the model with a solid line.
The model-data residuals are also shown.}
\label{decomp}
\end{figure*}

\subsection{The X-ray spectral analysis}
We present another X-ray spectral analysis of the NGC~7785 data. 
Thanks to the quality of the X-ray data and in view of the structure of
Eq.~(\ref{rybi}), we aim to deeply investigate the temperature profile 
of the X-ray gas emitting component of this isolated elliptical galaxy.

Source counts were extracted for the XMM-{\em Newton} EPIC-MOS1,-MOS2, and -pn from
two separate, but contiguous, regions: a 65\arcsec radius circle centered on the source,
and a concentric annulus with 65\arcsec and 130\arcsec radii. The background spectrum
was extracted from a circular source-free region of ~90\arcsec,
at 3.8\,arcmin distance from the target. 
The X-ray spectra were analyzed with the XSPEC package (version 12.6.0;
\citealt{arnaud}). The source counts were stacked into energy bins such that
each bin has a significance of at least 2$\sigma$, after background
subtraction. The quoted errors on the best-fit parameters correspond to the
90\% confidence level for one interesting parameter
(i.~e., $\Delta\chi^2 = 2.71$; \citealt{avni}).

In both cases the best-fit function of the source counts to fit the hot gas
includes a power-law with index fixed at $\Gamma$\,=2, which models the central
X-ray binaries or AGN component, plus a thermal ({\it mekal} model) component
with abundances fixed at 50\% the solar values. For both components the
appropriate Galactic hydrogen column density ($N_\mathrm{H}$) along the line of sight
\citep{dickey} has been taken into account. We found (see Fig.~\ref{7785spec}
and Table~\ref{7785tab}) $ kT_\mathrm{g} = 0.32^{+0.09}_{-0.05}$~keV in the outer annulus and $
kT_\mathrm{g} = 0.57^{+0.01}_{-0.02}$~keV in the inner circle, 
in agreement with \citet{memola}. Hereafter, we assume 
$T_\mathrm{mekal}= T_\mathrm{g}$.

A best-fit for the outer region ($\chi^{2}/\mathrm{d.o.f.} = 39/39$) with two free
parameters gives the $k T = 0.32^{+0.09}_{-0.05}$~keV value, which is 
incompatible with 
$k T = 0.56\pm0.02$~keV that refers to the hypotesis of constant temperature.
Our result is fairly robust: a constant value of $k T_\mathrm{g}$ out to
$130\ $arcsec can be excluded at $> 4\sigma$ level, and consequently a
temperature gradient emerges that we fit linearly. 

In detail, the data are well reproduced by
\begin {equation} 
k T_\mathrm{g}(r)=\left[0.6 - 1.2 \times 10^{-2} (r - 3)\right] \mathrm{keV},
\label{eq5}
\end {equation}
with $r$ in kpc. This result is not surprising and agrees with the
general behavior of isolated ellipticals: \citet{nagino} find that these
objects tend to show a negative temperature gradient. Notice that this
variation implies in Eq.~(\ref{rybi}) a mild contribution from the
temperature gradient term that it is smaller than the one from the gas
density gradient.  Therefore, the effect of the {\it uncertainty} in the
temperature profile on the mass distribution of Eq.~(\ref{mr}) is
negligible.

In NGC~7052 we cannot statistically support any evidence of gas temperature variation.
For this galaxy, then, $dT/dr =0$ as claimed in \citet{memola}.
A constant temperature has also been observed in some other isolated
ellipticals, e.g.~NGC~57, NGC~7796, IC~1531 \citep{osully07}.
%

\section {Structural parameters of the stellar spheroids} 
NGC~7785 is known to display boxy isophotes \citep{lauer}, and taking this into
account, we re-derive its structural parameters from the $K_\mathrm{s}$-band image
(see Fig.~\ref{decomp}) taken from the Two Micron All Sky Survey (2MASS) atlas. The
pixel size is 1 arcsec, and the point-spread function full-width half maximum is
$2.9$ arcsec. We used the two-dimensional galaxy fitting algorithm GALFIT
\citep{galfit} for fitting one S\'{e}rsic component, which is allowed to have disky
or boxy isophotes. The sky value is fixed to the value provided in the 2MASS image
header. The half-light radius is found to be $r_\mathrm{e}=22.70\arcsec$ and the
S\'{e}rsic index $n=3.8$, very close to the previously found de Vaucouleurs fit
$r_\mathrm{e}=19.5\arcsec$ \citep{bender}. The GALFIT boxiness parameter for the
best-fit model is $C_0=0.19$, and we verified that the addition of this parameter
in the case of this galaxy does not change the half-light radius or the S\`{e}rsic
index. If, instead of fixing the sky value, we let it be a free parameter, 
which is determined by GALFIT, both the $r_\mathrm{e}$ and $n$ values are 10\% lower. 
This effectively sets the uncertainty of the fit, which is negligible 
for the results in the succeeding sections of this work.
The performance of this fit can be seen from
the residual image in Fig.~\ref{decomp} (c). 

We also compared the surface brightness profile from the above fit
with the surface brightness profile of the galaxy image constructed
by means of the IRAF task ELLIPSE. The result is shown in Fig.~\ref{decomp}.

In NGC~7052 the photometry is simple and well studied by \cite{bender}: the distribution
of the stars is well described by the usual de Vaucouleurs spheroid with half
light radius of $r_\mathrm{e}=24\arcsec$. We confirmed this value with a two-dimensional
image decomposition of a 2MASS $K_\mathrm{s}$-band image performed as above.
However, we omit the plots for brevity reasons.

\section{Stellar spheroid mass in NGC~7052 and NGC~7785}
It is crucial to obtain an independent estimate of the mass of the stellar
spheroid for these objects. We computed them by fitting the observed
spectral energy distribution (SED) with synthetic spectra. We proceeded as follows: photometric data in
B,V, and 2MASS J,H,K$_\mathrm{s}$ bands were adopted from NED (homogenized NED
values) and corrected for Galactic extinction from \citet{schlegel}. 
We used the total magnitudes in 2MASS bands, which are obtained by 
integrating the radial surface brightness profile out to $r_\mathrm{tot}=87.1\arcsec$ 
and $82.2\arcsec$ for NGC~7052 and NGC~7785, respectively. Given that half-light 
radii for these two galaxies are $r_\mathrm{e}=24\arcsec$ and $22.7\arcsec$, integration out to nearly four times these radii provides a good estimate of the total galaxy magnitude \citep{jarrett}. The error estimate for the total magnitude is 0.02~mag for both galaxies.
The uncertainty introduced by the extinction correction and possible systematic differences between
bands is accounted for by considering an additional 0.05~mag error in quadrature.
Photometric data for NGC~7785 were available in the above five bands, while
for NGC~7052 the $V$-band data were unavailable.

\begin{figure}
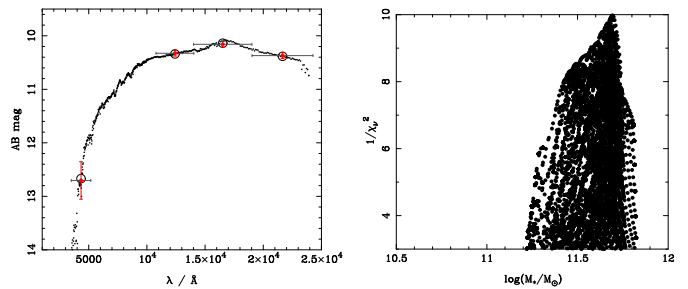

\resizebox{9cm}{!}{
\hspace*{1cm}
\psfig{figure=n7052-5pc-sed.ps,width=16.5cm,angle=-90}
\hspace*{1cm}
\psfig{figure=n7052-5pc-starmass.ps,width=16.5cm,angle=-90}
}
\caption{
NGC~7052. {\em Left:} Model fit to photometry 
(see color figure online). 
The best-fit spectral template is marked with 
points, the corresponding model magnitudes in 
the observed bands with circles. The observed 
magnitudes are indicated with filled red circles 
and error bars. The (black) bars in the wavelength
direction denote the bandwidth of respective filters.
{\em Right:} The likelihood of the stellar mass 
estimates.
}
\label{spectra7052}
\end{figure}

\begin{figure}
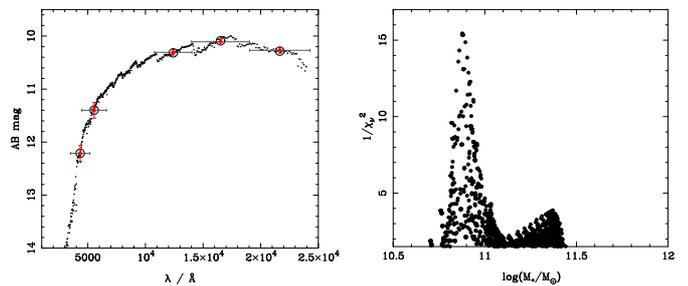

\resizebox{9cm}{!}{
\hspace*{1cm}
\psfig{figure=n7785-5pc-sed.ps,width=16.5cm,angle=-90}
\hspace*{1cm}
\psfig{figure=n7785-5pc-starmass.ps,width=16.5cm,angle=-90}
}
\caption{Same as Fig.~\ref{spectra7052} for NGC~7785. 
}
\label{spectra7785}
\end{figure}

The stellar mass is obtained from the 
mass-to-light ratio of the best-fitting model and the $K_s$ values (see Table 1). We fit the synthetic spectra to the observed spectral energy distribution by
using the modified version {\it hyperzmass} of the HyperZ code
\citep{bolzonella}, kindly provided by M. Bolzonella. The redshifts are fixed
to $z=0.0156$ and $z=0.0127$, for NGC~7052 and NGC~7785, respectively. The
adopted set of synthetic spectra consists of composite stellar populations with
two bursts of star formation, based on simple stellar population models of
\citet{m05} and constructed using the code {\em GALAXEV} \citep{bc03}. The
Salpeter initial mass function and solar metallicity are adopted. The strength
of the more recent burst is 5\% of that of the older one, and they are
separated by a period varying in the range 0-11 Gyr. The adopted star-formation
rate is consistent with what is typically found for elliptical galaxies and
includes an old major burst at approximately the halo-formation redshift and
the presence of a weaker burst in the last several gigayears. 

For NGC~7052 the best-fit model (Fig.~\ref{spectra7052}, filled black circles)
reproduces the data well (model magnitudes: open circles; data: filled red
circles and error bars). It consists of two bursts of star formation, one
11.3~Gyr ago, and the second one, of relative strength $5\%$, still occurring.
The resulting stellar spheroid spectrophotometric mass 
is $\log (M_{sph}^s/\Msun) = 11.6^{+0.1}_{-0.2}$, 
at 2$\sigma$ fitting uncertainty, then 
$\log (M_{sph}^s/L_B) = 0.84^{+0.1}_{-0.2}$.

For NGC~7785 the
best-fit model reproduces the data well, in this case it points to a main burst 
9.0~Gyr ago, followed by the second weak one 8.0~Gyr ago. For this object, a single
burst of star formation about 9.0~Gyr ago would also be a satisfactory model.
The stellar spheroid spectrophotometric mass is found to be 
$\log  (M_{sph}^s/\Msun) = 10.9^{+0.3}_{-0.1}$,
at 2$\sigma$  fitting uncertainty, then
$\log (M_{sph}^s/L_B) = 0.20 ^{+0.3}_{-0.1}$.

Figures \ref{spectra7052} and \ref{spectra7785} show the spectral template
fitting (left) and the mass determination (right) for NGC~7052 and NGC~7785,
respectively. In both cases a large number of different SFRs yield a stellar
mass similar to the one we assume. Models in the likelihood plot differ by 
the age of the two components and the reddening, which is for simplicity 
assumed to be the same for both components.
The mass estimates obtained by this method are
very robust: they primarily depend on the normalization of the SED in the
near-infrared bands, and are relatively insensitive to the details of the 
star-formation history \citep{maraston10}. In particular, changing the star-formation 
history to a single star-formation episode of varying length and functional form
results in a maximum 10\% change in the stellar mass (for the best-fit model) 
for NGC~7785 and 20\% for NGC~7052; varying metallicity in the range between 
half-solar and two solar metallicity leaves the mass unchanged for NGC~7785 
and 30\% smaller for NGC~7052; changing the relative strength of the second 
burst from 5\% to 10\% gives the same mass for NGC~7785 and a 10\% lower 
mass for NGC~7052. Again, these modifications mainly influence the 
implied age for the population, with a small effect on the deduced stellar mass, 
which is well within the quoted error bars. 
Using the Chabrier or Kroupa initial mass function instead of the Salpeter one will 
result in mass-to-light ratios $\approx 1.7$ times lower \citep{m05}.

\section{Mass models}

\begin{figure*}
\vskip -1truecm
\resizebox{18cm}{!}{
\hspace*{0.011cm}
\psfig{figure=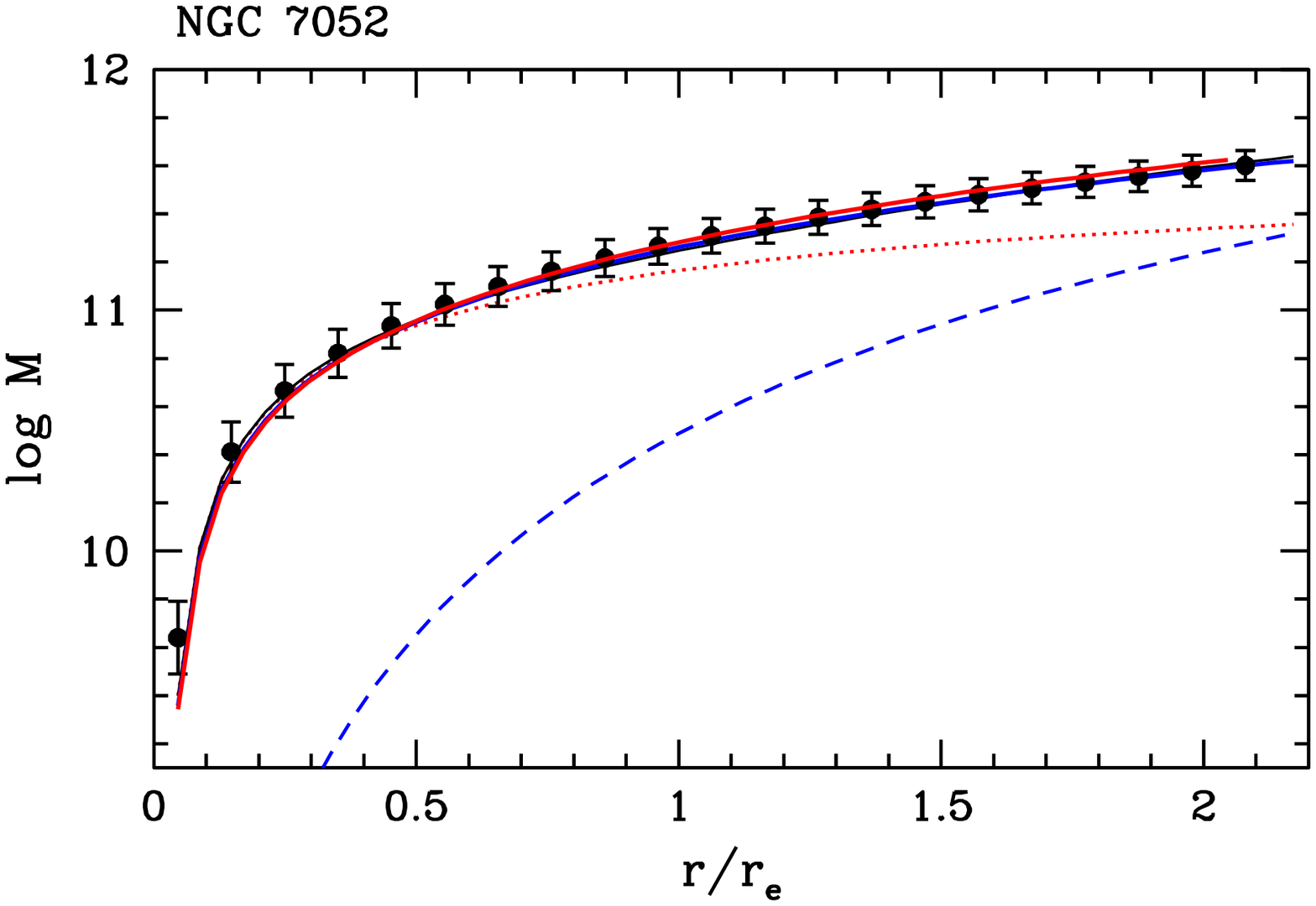,width=17.5cm} 
\hspace*{1cm}
\psfig{figure=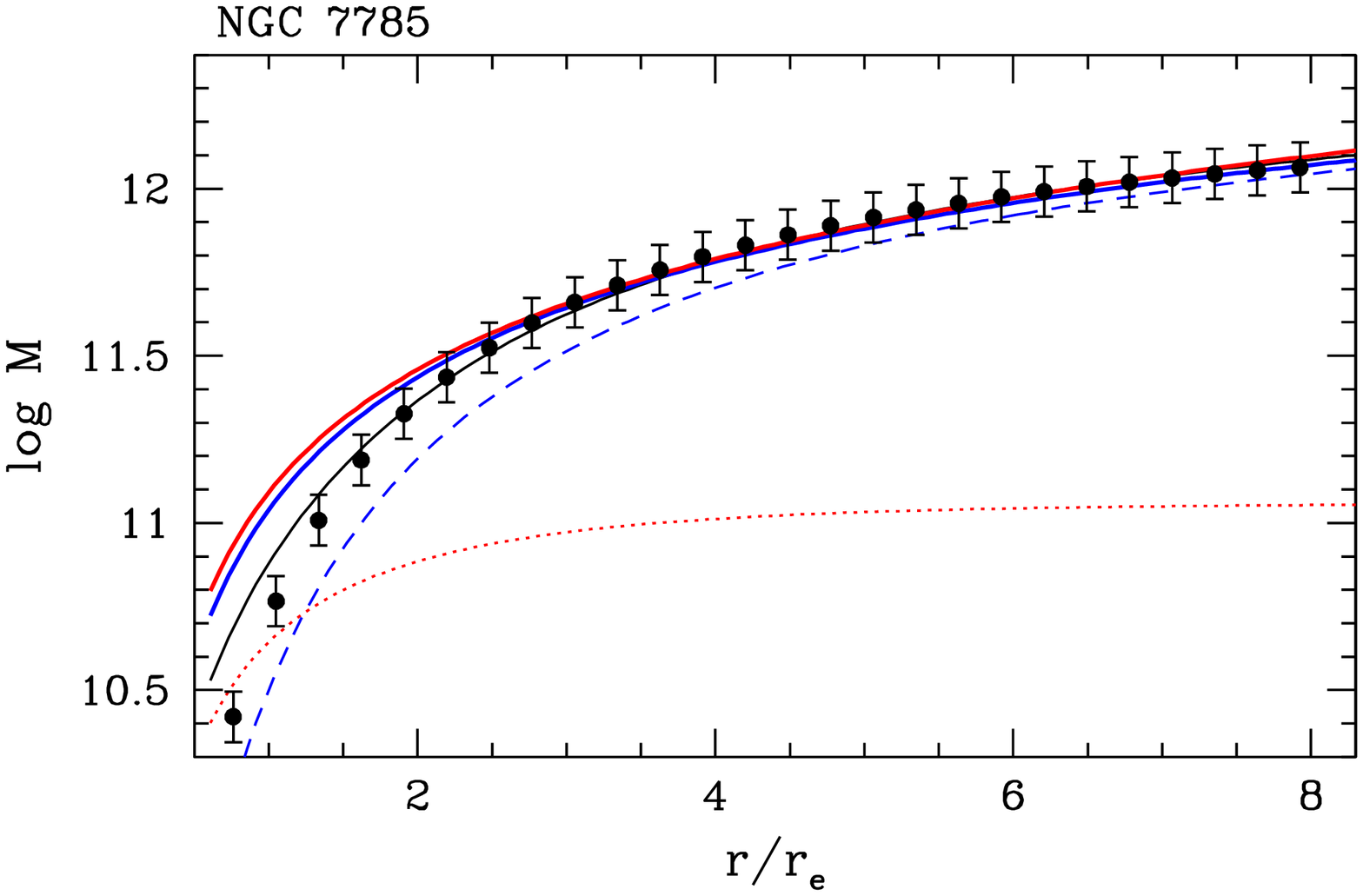,width=17.5cm} 
}
\caption {{\em Left:} Mass models of NGC 7052.  The X-ray derived dynamical mass 
(filled circles with error bars) is modeled (thin black
line) including a stellar spheroid (dotted line) and a 
Burkert halo (dashed line). Also shown is the best-fit solution for the  
total mass for models that assume a NFW profile to represent the dark 
halo (blue line) or the MOND framework (red solid lines). 
$r_e$ corresponds to 7.7 kpc.
{\em Right:} Mass models  of 
NGC 7785. $r_e$ corresponds to 5.6 kpc (see color figure online).}
\label{masfig} 
\end{figure*}

We fitted the profile of the gravitating mass $M(r)$ obtained by Eq.~(\ref{rybi})
with a mass model $M_\mathrm{model}(r)$ that includes a luminous spheroid and
a dark halo (Bertola et al.~1993; Kronawitter et al.~2000): 
\begin{equation}
M_\mathrm{model}(r)=M_\star(r)+M_\mathrm{h}(r).
\label{masmod}
\end{equation}
In these objects the stellar surface brightness follows a S\'{e}rsic
profile with index $n$ and effective radii
$r_\mathrm{e}$ given in Table~\ref{par-tab2}. Once we deproject $r_\mathrm{e}$ 
under the reasonable assumptions of spherical symmetry and constant stellar
mass-to-light ratio, we obtain the spatial density $\rho_\star(r)$ and mass profile
$M_\star(r)= \int_0^r 4 \pi r'^2 \rho_\star(r') dr'$ of the
stellar component. 

Let us note that in NGC~7052 the X-ray mass measurement extends (only)
to 16~kpc, corresponding to $ \simeq 2 r_\mathrm{e}$. Not surprisingly, in
this object the dynamical mass matches the spheroid mass over most of the
probed region, and the contribution of the DM to the gravitating matter
is small overall. Thus, to derive the structural parameters of the former
component from the latter leads to results with sizable uncertainties. 

We assume as in spirals that the dark matter is distributed in a spherical
halo with a Burkert density profile \citep{salucci00}:
\begin{equation} 
 \rho (r)={\rho_0\, r_0^3 \over (r+r_0)\,(r^2+r_0^2)}~,
\label{eq8}
\end{equation}
where the core radius $r_0$ and the central density $\rho_0$ are free
parameters. The dark mass profile $M_\mathrm{h}(<r)$ takes the form:
\begin{equation} 
 M_\mathrm{h}(r)= M_0 \Big\{ \ln \Big( 1 + \frac{r}{r_0} \Big) -
 \tan^{-1} \Big( \frac{r}{r_0} \Big) +{1\over {2}} \ln \Big[ 1
 +\Big(\frac{r}{r_0} \Big)^2 \Big] \Big\},
\label{eq9} 
\end{equation}
where $M_0\equiv 6.4 \rho_0 r_0^3$.

Ellipticals comply with the inner baryon dominance paradigm like spirals
\citep{albada85,salucci99}. 

The mass profile $M(r)$ obtained from the above equations,
with $M_\mathrm{model}\;(r, M_\star, \rho_0, r_0) $ given by Eq.~(\ref{masmod}) for
our two objects is fitted by minimizing the $\chi^2$. Estimated uncertainties are
$\delta\log M = 0.075\ $dex and 0.06~dex for NGC~7052 and NGC~7785,
respectively. The results are shown in Fig.~\ref{masfig}; the structural
galaxy parameters, i.e., the spheroid mass, the mass, the core radius, and the
central density of the halo are given in Table~\ref{par-tab2}. 

The mass model with a Burkert halo and a S\'{e}rsic spheroid gives a good fit
to the distribution of the gravitating mass of our galaxies: almost all data
are reproduced within the 1$\sigma$ error bars and $\chi^2_\mathrm{reduced}\sim
1$. The transition region between the region dominated by the stellar spheroid
and the region dominated by the dark matter occurs at $(2.0\pm 0.3) \
r_\mathrm{e}$. In both galaxies, we find that the stellar baryonic component
dominates the gravitational potential inside $r_\mathrm{e}$, in agreement with
\citet{bertola}, and, e.g., \citet{fukazawa}. At $r_\mathrm{e}$, the fraction
of the gravitating mass contained in the stellar spheroid amounts to $(76 \pm
5)\%$ in NGC~7052, and $(84^{+5}_{-15})\%$ in NGC~7785 where the lower spatial
resolution makes the estimate more difficult. Then, the $(M/L)$ profiles at
small radii match those expected from the stellar component alone.  

We find that the stellar spheroid 
masses are M$_{\rm sph}=1.1^{+0.4}_{-0.3} \times
10^{11} \Msun$ in NGC~7785, and M$_{\rm sph}=3.6^{+1.3}_{-1.0} \times 10^{11} \Msun$ 
in NGC~7052,
where the quoted uncertainties are the fitting ones. 
The stellar mass fraction inside $r_e$ found here for both galaxies is
quite consistent with earlier works (e.g.,~Gerhard et al.~2001; Thomas
et al.~2009, and several recent SLAC papers starting with Koopmans et
al.~2006).
For NGC 7052 one by-product of the good agreement we find between the 
X-ray mass profile and the stellar one
indicates that in this system there are
no strong deviations from the hydrostatic equilibrium condition,
which may lead to an underestimate of the cumulative mass,
in agreement with previous studies \citep{johnson}.

In any case, we find that in NGC~7052 the X-ray derived mass profile
is well fitted by the mass models. This is not obvious and in turn indicates that
these measurements are very reliable. Indeed, between $0.3r_e$ and $0.8r_e$
the mass profile of a de Vaucoulers spheroid has a unique distinct feature 
that we find in the data.

Remarkably, these values agree very well with the above SED fitting 
estimates. Mass modeling leads in the $B$-band to a spheroid mass-to-light ratio
of M$_{\rm sph}$/L$_{\rm B} = 6.3 \Msun/L_{\rm B_\odot}$ for NGC 7052, 
and M$_{\rm sph}$/L$_{\rm B} = 2.3 \Msun/L_{\rm B_\odot}$ for NGC 7785,
with a fitting uncertainty of 30\% (see Table\,3).
Note that the value for NGC~7785 is on the lower side for ellipticals, 
but it is confirmed by spectral fitting. 
In fact, in NGC 7785 the spectrophotometric estimate is
M$_{\rm sph}^s$/L$_{\rm B} = 1.7 \Msun/L_{\rm B_\odot}$
(M$_{\rm sph}^s=8\times10^{10}\Msun$), while in NGC 7052 it is
M$_{\rm sph}^s$/L$_{\rm B} = 6.9 \Msun/L_{\rm B_\odot}$
(M$_{\rm sph}^s=4\times 10^{11}\Msun$).

Before we discuss the DM properties, notice that in NGC~7052 
the dark matter dominates the mass distribution
only at the outermost radii, for which we have X-ray data ($1.5 r_\mathrm{e}<r<2.0
r_\mathrm{e}$), which makes
the estimate of its structural parameters uncertain. 

The dark matter halos show a cored density distribution. The core radii
and the corresponding central densities take in NGC~7052 and NGC~7785
the values of $22.3 /8.7\ $kpc and $1.3 \times 10^{-24}/1.0 \times
10^{-23}\ \mathrm{g\ cm}^{-3}$. However, these estimates are uncertain:
even considering the mass profiles we infer from X-ray fluxes as
error-free, the fitting uncertainties on these two parameters would be
at the level of 50\%.  Furthermore, unlike for spirals where the
dynamical mass is quite well determined from the rotation curve, here
there is a non-negligible uncertainty in the (derived) mass profile that
we fitted with the model $\delta d\log M(r)/d\log r\simeq 0.15$, which
contributes an additional uncertainty of $\sim 30\%$ on the derived
values of $r_0$ and $\rho_0$. The comparison of these values with those
in spirals of the same mass must await more and more precise mass
models; however, we can already claim that DM halo around ellipticals
seem to be denser and with a smaller core radius, but the
product $\rho_0 r_0$ is still, approximately, the same as the 
one found in smaller objects and for different Hubble types 
(Donato et al.~2009).

Notice that in NGC 7785, similarly to what it is found in (some) other ellipticals, 
the X-ray derived mass profile in the very inner galaxy region, 
$r < r_e < 6$~kpc, corresponding to the innermost 2-3 data points, 
cannot be reproduced by any reasonable mass profile, including a Sersic spheroid, a 
standard DM halo or a combination of both. The dynamical X-ray mass 
appears, for $r \rightarrow 0$, to be progressively smaller than 
the actual data. This discrepancy, probably caused by the 
contribution to the X-ray emission of i) non-thermal pressures or ii) 
multiple-temperature components (see Das et al.~2010)
does not affect the main results of the mass modeling, however.
Indeed in the Burkert class spheroidal (BS) model
the spheroid mass has been obtained by imposing 
that at $2 r_e$, i.e., at a radius outside the troubled region, the 
spheroid and the dark halo masses equal the dynamical mass. 

Remarkably, we note that the resulting value for the spheroid
mass is consistent with the photometric estimates. 

Finally, we checked the effects on the resulting best fitting 
values of the DM halo structural parameters ($\rho_0$ and $r_0$) of 
1) a variation of the spheroid mass of $\pm30\%$; 2) by considering
the latter quantity as a free parameter; 3) including the two innermost data points. 
In all these checks, we found that the BS model fits
the dynamical mass very well and the resulting values of $\rho_0$ and $r_0$ 
lie always within 30\% those given in Table\,3.
The presence of a cored DM mass distribution relies on the fact that 
between $r_e$ and $3 r_e$, i.e., well outside the troubled region, the 
dynamical mass shows an increase steeper than linear. 

We can estimate the virial masses and radii for NGC~7052 and NGC~7785 by
extrapolating the mass models: $M_\mathrm{vir}\simeq 8.6 \times
10^{12} \Msun $, $ M_\mathrm{vir} \simeq 5.2 \times 10^{12} \Msun$,
$R_\mathrm{vir} = 476$~kpc, and $R_\mathrm{vir} = 405$~kpc.  
Indeed, in the present case the extrapolation of the outermost X-ray 
mass $M(R_o)$ to the virial radius is feasible. The X-ray mass
profiles indicate that the present data have reached, contrary to the 
"extended " available HI RCs in spirals, the region in which the circular 
velocity $\propto (M/R)^{0.5}$ has a maximum (Salucci et al.~2007). 

We are allowed to extrapolate $M(r)$
because we know from weak-lensing measurements that
DM halos around isolated galaxies are not truncated out to their
virial radii (Mandelbaum et al.~2006) and that their outer profiles are 
consistent with a Burkert/NFW profile. The extrapolation uncertainties 
are at the level of 15\% for NGC 7785, and 20\% for
NGC 7052. The uncertainty on $M(R_o)\propto \beta T(R_o) \beta 
r_c^{-1} $ is also small, because the observational uncertainties
on $T(R_o), \beta, r_c $ are at a level of 10\% (Memola et al.~2009). 

These estimates of the virial mass from direct mass measurements are
solid: these are two of the few cases in which this has been done so far.

This confirms that big ellipticals live in
halos two to four times more massive than those around big spirals.

The baryon fractions, i.e., the ratios of stellar and virial mass, are $0.05\pm
0.02$ and $0.03\pm 0.01$, where the main uncertainty comes from how we
extrapolate the dark mass profile out to the virial radius. The values
found agree with the determinations of \citet{shankar}
obtained by correlating the baryonic mass function with the halo mass
function. 

In summary, the fractional amount of the stellar component looks 
(within a factor of
two) similar to that found in the biggest spirals \citep{salucci07} and
is in any case much smaller than the cosmological ratio of $\sim 1/7$.
Given the large gravitational potential well of these galaxies, it
remains to be understood where 70\%- 80\% of the original cosmological
baryons went. The dark matter mass profile in the central parts
indicates a cored density distribution, although the estimate of the
characterizing structural parameters (central density and core radius)
is uncertain. 

\subsection {NFW halo profile}
We further investigated the halo mass profile by fitting the dynamical
mass profiles with the NFW halo profile that emerged in the $\Lambda$CDM scenario
(Navarro et al.~1997): 
\begin{equation} 
  M_\mathrm{NFW}(r)=
  M_\mathrm{vir} {\left[\log( 1+r/r_\mathrm{s})-
  (r/r_\mathrm{s})/(1+r/r_\mathrm{s})\right]
  \over{\log(1+c)-c/(1+c)}},
\label{mnfw}
\end{equation}
with $r_\mathrm{s}\equiv R_\mathrm{vir}/c$. 

We tested this mass distribution, leaving $c$ and $M_\mathrm{vir}$ as free
parameters, which is necessary to obtain an acceptable value for the reduced
$\chi^2_\mathrm{red.}$. For NGC~7785
we leave the value of the stellar mass unchanged with respect to the one
obtained above, because it is irrelevant for the total
gravitational potential; for NGC~7052 we leave it free.

The model fits (see the dashed line in Fig.~\ref{masfig}) 
are very satisfactory for NGC~7052
($\chi^2_\mathrm{red}<1$) and less acceptable for NGC~7785
($\chi^2_\mathrm{red}=3.6$).  
The resulting NFWS spheroid masses are reasonable and similar to
the BS ones. In particular,
in NGC~7052 the spheroid mass is $85\%$ of the one obtained for the Burkert
halo mass model.

Moreover, in the latter case the resulting
value for the concentration parameter $c=33$ is inconsistent with the
$\Lambda$CDM predictions of $c=8-9$ for objects $\sim 5\times 10^{12}\Msun$.  
This is not due to the mass modeling uncertainties, because the
gravitating mass inside 30~kpc $>10^{12} \Msun$ 
is much larger than the spheroid mass, therefore
coincides with the halo mass, which implies that this object displays an
intrinsically cored mass distribution. 

As a counter proof, 
the standard NFW DM
halo of  $10^{12} \Msun$ has $c=9$ 
(e.g.,~\citealt{bertola,kronawitter,klypin10,tortora2})
and it is completely inconsistent with the data. 

Note that halo models with $c \simeq 3$, which is equally inconsistent with
$\Lambda$CDM prediction, but in line with the results of
\citet{fukazawa}, have a $\chi^2_\mathrm{red}$ similar to cored best-fit
values.

Changing the values of the spheroid masses or considering the adiabatic
contraction process for both galaxies for the model with the NFW
halo and S\'{e}rsic spheroid (NFWS) gives a {\it worse} fit. 
The NFWS model vs observation discrepancy is very different with respect 
to the BS one. Firstly, it extends outside the (possibly) troubled region, 
out to $r\leq 2.5 r_e \leq 15$~kpc. Secondly, the discrepancy is also found 
in a region where only one component is present, the dark one, which 
completely dominates the potential with a known density profile. 
Furthermore, the (very high) value of $c$ does not depend on the 
innermost values of the dynamical mass, but on the fact that from $r 
\sim r_c\simeq 1.5 \ r_e \sim 8$~kpc out to the last measured point at 
$\sim 45$~kpc the dynamical mass (completely dominated by the dark 
component) increases perfectly linearly with the radius. 
Independently on how we treat the first two data at $r<r_e$, 
the NFWS mass model for NGC 7785 must have $c\simeq 35$.

\subsection{MOND}
We realize that our objects are in the Mondian regime for $r>5$~kpc.
Our two objects are  then well suited for testing the MOND paradigm, in
particular the case in which it is seen as a modification of the inertia
law, able to account for the ``missing light'' in galaxies. In this
framework the relation between the density of the gravitating matter and
its relative acceleration, when the latter is $\leq a_0 =1.3 \times
10^{-8} \mathrm{cm\ s}^{-2}$, is very different from the
Galilean-Newtonian case. As a consequence, for a baryonic mass
distribution $ M_\mathrm{b}(r)$, we measure (e.g. by applying the
hydrostatic equilibrium condition) a different/larger dynamical mass
$M_\mathrm{MOND}(r)$: 
\begin{equation}
M_\mathrm{MOND}(r)={1\over 2} M_\mathrm{b}(r) \left\{ 1+ \left[ 1+ {4.1 \times
10^9 \Msun \over M_\mathrm{b}(r) } \left(\frac{r}{1 \mathrm{kpc}}\right)^2
\right]^{0.5} \right\}.
\label{mondeq}
\end{equation}

Therefore, we test MOND by fitting $M(r)$ given by Eq. (\ref{masmod})
with $M_\mathrm{MOND}(r)$, leaving the mass of the spheroid as a free
parameter. We can anticipate that MOND will have difficulties accounting
for the DM of early-type systems.  They have just one baryonic
component, the S\'{e}rsic spheroid, whose mass converges well inside the
region for which we have the dynamical mass. This creates a mass
discrepancy that steadily increases with radius as a power-law. In
spirals MOND must account for a very similar behavior of the dynamical
mass but it has two additional advantages: 1) there are two baryonic
components -- the stellar and the HI disk, and 2) the mass of the latter
does not converge inside the region mapped by dynamical data. 

MOND shows a tension in NGC 7052, and fails the test for NGC~7785
(see Fig.~\ref{masfig}). 
The value $\chi_\mathrm{red}^2 \simeq 2$ of this mass model is fairly high.
However, even if we consider the MOND mass model satisfactory, 
the resulting best-fit values for the spheroid mass-to-light ratios
($M/L_\mathrm{B}\simeq 7$ and 15) are inconsistent with the colors of
these galaxies. 

\begin{table*}
\caption{ Structural properties. 
Columns:
(1) galaxy name; 
(2) last measured point, in arcsec;
(3) half-light radius in kpc; 
(4) S\'{e}rsic index;
(5) mass inside the last measured point, in units of $10^{11}\Msun $;
(6) spheroid mass in units of $10^{11}\Msun $; 
(7) mass-to-light ratio of the spheroid in $B$-band, in units of $\Msun/L_{\rm B_\odot}$;
(8) dark matter Burkert halo core radius in kpc; 
(9) dark matter Burkert halo central density in units of $10^{-24}\mathrm{g\ cm}^{-3}$;
(10) Burkert halo virial mass in units of $10^{12}\Msun$ ;
(11) concentration parameter $c$ of NFW halo; 
(12) NFW halo virial mass in units of $10^{12}\Msun$;
(13) Burkert halo virial radius in kpc;
(14) baryonic fraction.
}
\label{par-tab2}
\begin{center}
\begin{tabular}{cccccccccccccc}
\hline 
\hline
\noalign{\smallskip}
Source & lpt & r$_{\rm e}$ & n & M$_{\rm lpt}$ & M$_{\rm sph}$ & M$_{\rm sph}$/L$_{\rm B}$ & r$_{0}$ & $\rho_{0}$ & M$_{\rm vir}$ & $c$ &M$_{\rm v_{NFW}}$ & $R_\mathrm{vir}$ & $f_\mathrm{baryon}$ \\

(1)& (2)& (3)& (4)& (5)& (6) & (7) &(8) & (9)& (10)& (11)& (12) & (13) & (14) \\
\hline
\noalign{\smallskip}
NGC~7052 & 16.1 & 7.7 & 4   & 4.0 & 3.6 & 6.3 & 22.3 & 1.3 & 7.4 & 10 & 8.6 & 476 & 0.05 \\
NGC~7785 & 32.1 & 5.6 & 3.8 & 9.0 & 1.1 & 2.3 & 8.7 & 10.2 & 3.6 & 33 & 5.2 & 405 & 0.03 \\
\noalign{\smallskip}
\hline
\end{tabular}
\end{center}
\end{table*}

\section{Conclusions} 
We study the mass distribution of two isolated ellipticals by means of
their X-ray emission and stellar photometry. We find, and this is one of
the few cases at present for early-type galaxies in such suitable
environment, that a dark component must be present for $r>r_\mathrm{e}$,
because $r_\mathrm{e}$ is the half-light radius of the stellar spheroid.
More specifically, we find that inside $r_\mathrm{e}$ the stellar
component accounts for $\la 70\%$ of the gravitating mass. This
result is not new (see, e.g.,~\citealt{kronawitter,nagino}, where 22
objects with XMM-Newton and Chandra X-ray luminosities are mass-modeled). 
However, our finding is at odds with the claim of substantial
proportions of DM inside $r_\mathrm{e}$ \citep{humphrey} and with the
idea that the well known tilt of the fundamental plane of ellipticals is
due to systematic variations of the DM fraction inside $r_\mathrm{e}$
with galaxy properties. Indeed, our (and other) mass models of
ellipticals imply that the central stellar line-of-sight velocity
dispersion has very little contribution from the dark component and
essentially weighs only the stellar spheroid, whose structural
properties (mass-to-light ratio and half light radius)
therefore must be considered responsible for the observed tilt.

In our two objects the total mass does not increase with radius as a
(unique) power-law, as suggested by \citet{humphrey10}: $d\log M/d\log r$
decreases from $\sim 2$ in the innermost regions mapped by X-ray
emission ($\sim 0.3 r_\mathrm{e}$) to $\sim 1$ in the outermost
regions ($\sim 2 r_\mathrm{e}$), a behavior often found in the mass
structure of ellipticals \citep{kronawitter,fukazawa}.  The mass
profiles of  NGC~7052 and NGC~7785 are easily and naturally reproduced by
a combination of a S\'{e}rsic spheroid and a Burkert halo, where the
former is much denser and the latter is much more massive. The fits are
good (unfortunately) for a large range of values of the structural
parameters, i.e., the central DM density, the core radius, and the
spheroid mass. This agrees with previous results: ellipticals show
a variety of mass profiles, each of them reproducible by a quite large
variety of different mass models. Thus, in our objects we find no
evidence for the claim that the mass distribution in ellipticals is
subject to a ``cosmic conspiracy'', i.e., that all of them would show the
same (unexplained) profile, implying just a relationship rather
than the claimed fine-tuning in the DM and
luminous matter structural parameters. 

As in spirals, NFWS mass models 
have some difficulties reproducing the data: the best-fitting
values of the $\chi^2$, the mass to light ratio of the luminous
component, the halo concentration parameter are out of the expected
range. However, in ellipticals, a NFW profile is not necessarily the
correct present-day profile for a $\Lambda$CDM halo as well. The baryons
largely dominate the inner potential and their infall/contraction may
have modified the DM distribution \citep[e.g.,][]{klypin}.

Finally, our mass determinations are in tension with
the baryonic Tully-Fisher relation \citep{mcgaugh}, according to which in cosmic
structures from clusters to dwarf spheroids (dSphs), the circular
velocity $V=G M(r)/r^{0.5} $ and the baryonic mass $M_\mathrm{b}$ scale
as $\log M_\mathrm{b}= 4\log V +1.61$. In our objects we obtain a
reliable estimate of the stellar mass that coincides with the baryonic
one. Moreover, for our objects the mass profile does increase linearly
with radius in the outer regions, so that the ``circular velocity'' is
radially constant with radius and can be easily extrapolated 
out to its virial value. 
We find that for NGC~7052 ($V, M_\mathrm{b}$) =( $330\
\mathrm{km\ s^{-1}}, 3.6\times 10^{11} \Msun$), while for 
NGC~7785 ($V, M_\mathrm{b}$) = ($380 \ \mathrm{km\ s^{-1}}, 1.1
\times 10^{11} \Msun$). 
These objects have baryonic masses that fall short of the predicted value from
this relation by a factor $\sim$2 and $\sim$8.

By investigating  two isolated ellipticals 
by means of their X-ray and 
optical emissions we find that
1)  they have  a considerable amount 
of DM, but only  at radii $> r_e$;  
2) their  DM halos  have a shallower  
distribution than that emerging in the $\Lambda$CDM scenario; 
3) these objects are in the Mondian regime, but do not agree with 
its predictions; 
4) these objects do not comply with the baryonic Tully Fisher.

\begin{acknowledgements}
We would like to thank C.~Maraston and J.~Pforr for providing the SED
templates and for helpful discussions. 
We also thank E.~Pompei, and an anonymous referee for  
comments, which improved the manuscript.
E.M.~thanks L.~Ballo, G.~Trinchieri, and A.~Wolter for priceless scientific hints. 
{\em This research was performed despite of the financial policy by the
Italian government.}
\end{acknowledgements}

\end{document}